\documentclass[]{spie}  

\newcommand{\Planck}{{\it Planck}}

\usepackage{amsmath,amsfonts,amssymb}
\usepackage{graphicx}
\usepackage[colorlinks=true, allcolors=blue]{hyperref}
\usepackage{color}

\makeatletter
\newcommand\footnoteref[1]{\protected@xdef\@thefnmark{\ref{#1}}\@footnotemark}
\makeatother

\title{BFORE: A CMB Balloon Payload to Measure Reionization, Neutrino Mass, and Cosmic Inflation} 

\author[ ]{Sean Bryan$^\mathrm{a}$, Peter Ade$^\mathrm{b}$, J. Richard Bond$^\mathrm{c}$, Francois Boulanger$^\mathrm{d}$, Mark Devlin$^\mathrm{e}$, Simon Doyle$^\mathrm{b}$, Jeffrey Filippini$^\mathrm{f}$, Laura Fissell$^\mathrm{g}$, Christopher Groppi$^\mathrm{h}$, Gilbert Holder$^\mathrm{f}$, Johannes Hubmayr$^\mathrm{i}$, Philip Mauskopf$^\mathrm{h}$, Jeffrey McMahon$^\mathrm{j}$, Johanna Nagy$^\mathrm{k}$, C. Barth Netterfield$^\mathrm{l}$, Michael Niemack$^\mathrm{m}$, Giles Novak$^\mathrm{n}$, Enzo Pascale$^\mathrm{o}$, Giampaolo Pisano$^\mathrm{b}$, John Ruhl$^\mathrm{p}$, Douglas Scott$^\mathrm{q}$, Juan Soler$^\mathrm{r}$, Carole Tucker$^\mathrm{b}$, and Joaquin Vieira$^\mathrm{f}$}
\affil[a]{School of Electrical, Computer, and Energy Engineering, Arizona State University, Tempe, AZ, USA}
\affil[b]{School of Physics and Astronomy, Cardiff University, Cardiff, UK}
\affil[c]{Canadian Institute for Theoretical Astrophysics, University of Toronto, Toronto, ON, Canada}
\affil[d]{Institut d'Astrophysique Spatiale, Orsay, France}
\affil[e]{Department of Physics and Astronomy, University of Pennsylvania, Philadelphia, PA, USA}
\affil[f]{Department of Physics, University of Illinois at Urbana-Champaign, Urbana, IL, USA}
\affil[g]{ational Radio Astronomy Observatory, Charlottesville, NC, USA}
\affil[h]{School of Earth and Space Exploration, Arizona State University, Tempe, AZ, USA}
\affil[i]{National Institute of Standards and Technology, Boulder, CO, USA}
\affil[j]{Department of Physics, University of Michigan, Ann Arbor, MI, USA}
\affil[k]{Dunlap Institute for Astronomy and Astrophysics, University of Toronto, Toronto, Canada}
\affil[l]{Department of Astronomy and Astrophysics, University of Toronto, Toronto, ON, Canada}
\affil[m]{Department of Physics, Cornell University, Ithaca, NY, USA}
\affil[n]{Department of Physics and Astronomy, Northwestern University, Evanston, IL, USA}
\affil[o]{Department of Physics, Sapienza Universit\`{a} di Roma, Rome, Italy}
\affil[p]{Department of Physics, Case Western Reserve University, Cleveland, OH, USA}
\affil[q]{Department of Physics and Astronomy, University of British Columbia, Vancouver, Canada}
\affil[r]{Max Planck Institute for Astronomy, Heidelberg, Germany}

\authorinfo{Send correspondence to S.A.B.: sean.a.bryan@asu.edu}
\begin{document}
\maketitle

\begin{abstract}
BFORE is a high-altitude ultra-long-duration balloon mission to map the cosmic microwave background (CMB). During a 28-day mid-latitude flight launched from Wanaka, New Zealand, the instrument will map half the sky to improve measurements of the optical depth to reionization tau. This will break parameter degeneracies needed to detect neutrino mass. BFORE will also hunt for the gravitational wave B-mode signal, and map Galactic dust foregrounds. The mission will be the first near-space use of TES/mSQUID multichroic detectors (150/217 GHz and 280/353 GHz bands) with low-power readout electronics.
\end{abstract}



\section{INTRODUCTION}
\label{sec:intro}  

The scientific potential of precision CMB polarization measurements has inspired a thriving field of programs to detect and characterize these signals from the ground, balloons, and space. The faint $B$-mode polarization of the CMB is of particular interest. Among a wide range of key science goals \cite{CMBS4-ScienceBook,CMBS4-CDT-Report}, ground-based CMB experiments aim to measure $B$-modes at arcminute angular scales
to measure neutrino mass and the history of structure formation. 
Detecting the neutrino mass with the CMB also requires improved CMB polarization data at large angular scales to measure the optical depth to reionization $\tau$, which may be difficult to access from the ground. Ground-based CMB experiments also aim to probe inflationary physics by measuring $B$-modes at degree angular scales. Achieving all key CMB science goals also requires improved foreground data, especially dust data at higher frequencies and arcminute to large angular scales which are difficult to access from the ground.

BFORE is a balloon-borne millimeter-wave cosmic microwave background (CMB) polarimeter designed to be flown on a NASA ultra-long duration balloon (ULDB) launched from Wanaka, New Zealand. The mission is designed to take advantage of the unique benefits of the ULDB platform: access to high frequencies (needed for galactic dust characterization), access to the largest angular scales (needed to measure reionization), and long integration time (needed for sensitivity and control of systematics). BFORE aims to: \textit{(i)} precisely measure the optical depth of reionization $\tau$, breaking parameter degeneracies needed to detect neutrino mass with the CMB; \textit{(ii)} measure or set an upper limit on the gravitational wave signal from cosmic inflation both at degree-scales and separately at large angular scales; \textit{(iii)} make a half-sky map of CMB dust foregrounds at $\sim$ few arcminute resolution with legacy value; and \textit{(iv)} make a deep-field map to measure galaxy cluster properties and star formation. The capabilities of mid-latitude ULDB allow BFORE to target all four science goals in a single flight.

\begin{figure}
\begin{center}
\begin{tabular}{c}
\includegraphics[angle=0,width=1.0\textwidth]{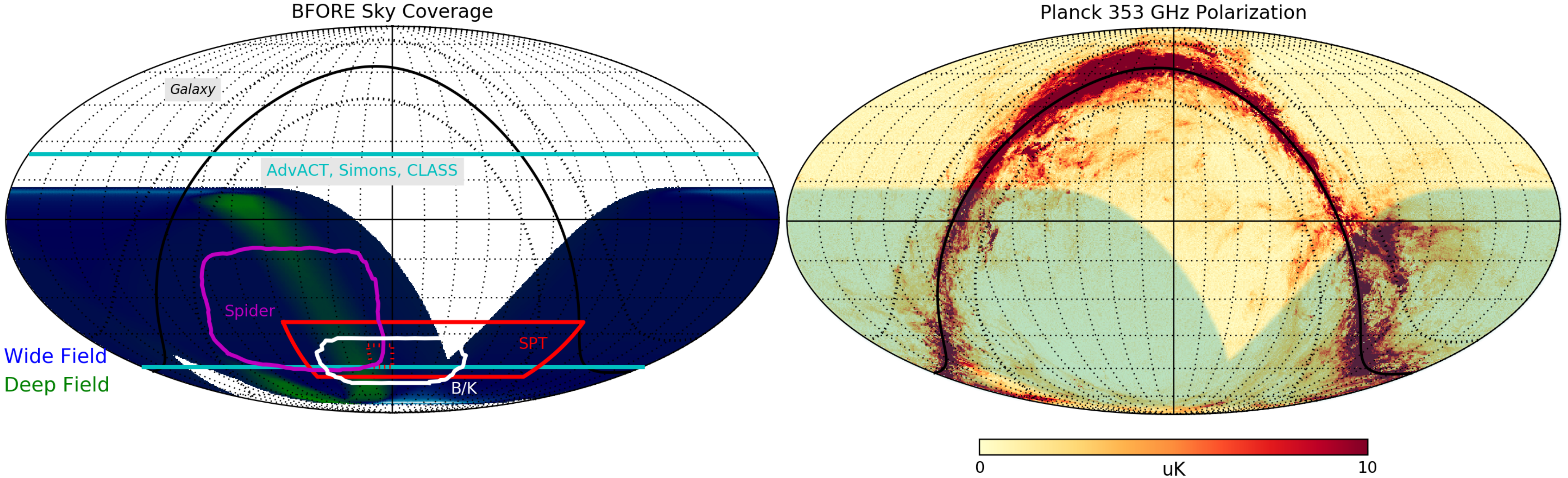}
\end{tabular}
\end{center}
\caption{ 
 Simulated sky coverage for a 28-day BFORE ULDB flight launched from New Zealand on April 15th, 2022. The wide field, shown in blue in the left panel, is observed by spinning the instrument at 6 deg/s throughout every night of the flight. This wide field covers $50\%$ of the sky and uses the unique sky coverage provided by mid-latitude ULDB to overlap multiple complementary ground-based CMB experiments including Simons Observatory and Simons Array \cite{arnold/etal:2014}, AdvACT \cite{henderson/etal:2015}, CLASS \cite{essinger-hileman/etal:2014}, SPT-3G \cite{benson/etal:2014}, and BICEP/Keck \cite{bicep2}. Our sky coverage also overlaps the Spider balloon instrument \cite{fraisse13}. The right panel shows the polarization map (i.e. $\sqrt{Q^2 + U^2}$) measured by \Planck\ at 353\,GHz, with the BFORE-Wide field highlighted. BFORE will spend the daytime portion of its flight scanning a 2500\,deg$^2$ deep field. The deep field hits map is shown in green in the left panel, and is in a low-foreground portion of the southern sky.
\label{fig:sky}}
\end{figure} 

In addition to achieving its own scientific goals, BFORE will complement current and future ground-based CMB surveys, providing maps
to the broader community with low noise, high fidelity over a large range of frequencies and angular scales, and overlapping sky coverage with ground-based experiments from both Chile and Antarctica. BFORE uses the mid-latitude ULDB platform to overlap with the sky coverage of both sites, and to reconstruct large angular scale $E$-modes needed to measure $\tau$, both of which are critical to the legacy value of the BFORE data. In contrast, conventional Antarctic ballooning has limited access to the sky coverage and large angular scales needed to measure $\tau$, and also has limited sky coverage overlap with Chile.

An overview of the BFORE mission is presented here \cite{bryan16}. In this paper, we present an update of the current status of the mission, including improved forecasting of the science reach of BFORE, additional information about the cryostat and payload weight budget, as well as new developments in commercial low power detector readout electronics.

\section{Science with BFORE}

BFORE will use ULDB plaform to take data at high frequencies and large angular scales (150, 217, 280, and 350 GHz, across half of the sky, at 2.6+ arcminute angular resolution). This will be a rich dataset answering several important science questions. The large angular scale E-mode data will measure the optical depth to reionization $\tau$. This will break parameter degeneracies and will enable a detection of neutrino mass with the CMB. The BFORE data will detect or set a limit on the inflationary gravitational wave signal at degree-scales and large angular scales, each at the $r \sim 0.01$ level. Measuring the large angular scale modes provides a limit on $r$ that is independent from degree-scale $r$ limits, and also constrains the spectral index $n_T$ of the gravitational wave signal. BFORE will provide foreground cleaning data to ground based CMB lensing measurements, and also will measure galaxy cluster properties. These science goals for BFORE are reviewed in more detail elsewhere \cite{bryan16}, so here we discuss improved forecasting for the $\tau$ and neutrino mass science goal, verifying that our MCMC-based forecasting tool agrees with the independently-developed CMB4CAST\cite{errard16} tool, showing that BFORE is robust to foregrounds, and showing that achieving the $\tau$ science goal has margin on detector sensitivity and stability.

\begin{table}[]
\caption{BFORE Telescope and Receiver Parameters, and Map Depth Forecast} 
\label{tab:inst}
\begin{center}       
\begin{tabular}{| l l l | }\hline \hline 
\textbf{Telescope:} &Temperature&250\,K (Primary), 4.2\,K (Secondary)\\ 
& Primary diameter&1.35\,m, emissivity $\le 0.005$\\ \hline
\textbf{Detectors:} 

& Central frequencies  & {\bf 150~~217~~~280~~\hspace{0.1em}353~~\hspace{0.1em}GHz}\\

\textit{($\mathit{3.4}^\circ$-diameter FOV)} & Number of TESs & 2400~~2400~~2880~~2880 \\ 

\textit{(100 mK fridge)}	 & Detector NEP & 6.1~~~~9.2~~~~\hspace{0.1em}9.6~~~ 16~~~~~\hspace{0.1em}{\rm aW}/${\sqrt{{\rm Hz}}}$ \\ 

\textit{(40\% end-to-end} & Sky+atmosphere+optics loading~~ & 0.9~~~~1.5~~~~1.3~~~~\hspace{0.07em}2.8~~~~\hspace{0.1em}pW\\ 

\textit{optical efficiency)} & Background NEP~~ & 15~~~~~23~~~~~24~~~~\hspace{0.18em}41~~~~~\hspace{0.1em}{\rm aW}/${\sqrt{{\rm Hz}}}$\\
 & Single-detector Sensitivity (CMB)~~ & 103~~~164~~~~389~~~\hspace{0.07em}703~~~~$\mu$K${\sqrt{{\rm s}}}$\\
  & Single-detector Sensitivity (RJ)~~ & 60~~~~~57~~~~~71~~~~\hspace{0.29em}62~~~~~$\mu$K${\sqrt{{\rm s}}}$\\ \hline

\textbf{Half-sky map:}  & Map Depth (CMB) & \textbf{19.2}~~\textbf{30.5}~~65.9~~\hspace{0.1em}119~~~$\mu$K-arcmin\\
\textit{(28 nights} & Map Depth, Dust scaled to 150\,GHz & 19.2~~~8.96~~\hspace{0.15em}\textbf{8.48}~~\textbf{6.78}~$\mu$K-arcmin\\
\textit{w/ 90\% obs. eff.)}& Map Depth (RJ) & 11.2~~~10.6~~\hspace{0.05em}12.1~~\hspace{0.22em}10.5~~$\mu$K-arcmin \\ \hline
\textbf{Deep map:} & Map Depth (CMB) & \textbf{6.79}~~\textbf{10.8}~~23.3~~\hspace{0.13em}42.1~~$\mu$K-arcmin\\
\textit{(28 days }& Map Depth, Dust scaled to 150\,GHz & 6.79~~~3.17~~\hspace{0.13em}\textbf{3.00}~~\textbf{2.40}~$\mu$K-arcmin\\
\textit{w/ 90\% obs. eff.)}& Map Depth (RJ) & 3.95~~~3.73~~\hspace{0.145em}4.28~~\hspace{0.23em}3.70~~$\mu$K-arcmin \\ \hline
\textbf{Beam:} & FWHM & 6.1~~~~4.2~~~~\hspace{0.18em}3.3~~~~\textbf{2.6}~~~arc-minutes \\ 
&Bandwidth & 38~~~~~60~~~~~50~~~~~90~~~~GHz \\ 

\hline 
 \end{tabular}
\end{center}
\end{table} 

\subsection{Optical Depth and Neutrino Mass}

\begin{figure}
\begin{center}
\begin{tabular}{c}
\includegraphics[width=1.0\textwidth]{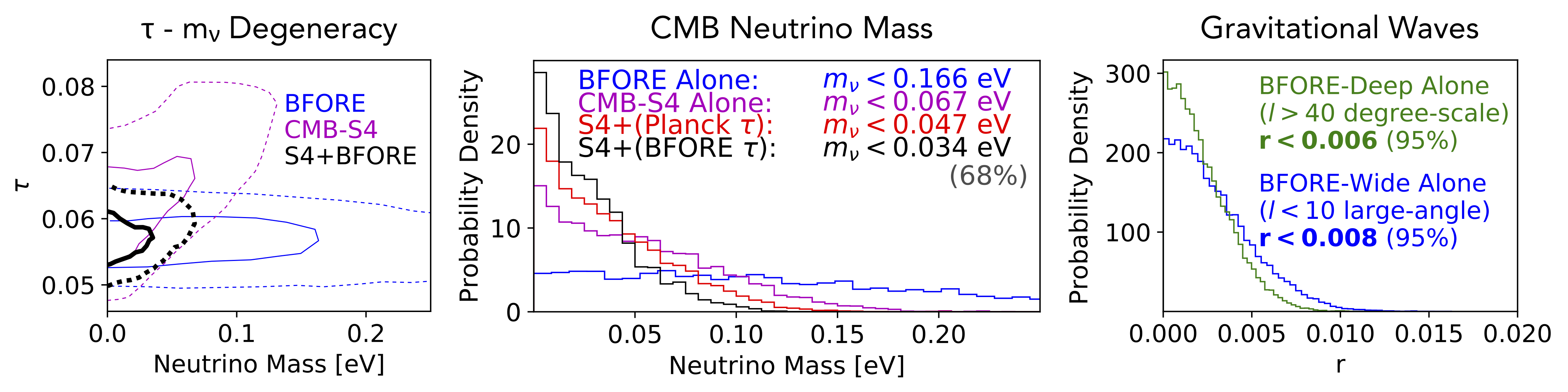}
\end{tabular}
\end{center}
\caption{\textbf{Left:} Ground-based CMB measurements (lacking sensitivity to large angular scales due to atmospheric fluctuations) suffer from a degeneracy between $\tau$ and neutrino mass. BFORE-Wide breaks this degeneracy with $E$-mode measurements at large angular scales. \textbf{Center: } BFORE-Deep measures neutrino mass independently, and the improved BFORE $\tau$ measurement breaks degeneracies to significantly improve the neutrino mass sensitivity of Simons Observatory or the proposed CMB-S4 experiment. Forecasting done by the CMB-S4 team for the Science Book \cite{CMBS4-ScienceBook} shows that Simons Observatory and/or CMB-S4, $\tau$ constraints better than \Planck's but at the level BFORE will provide, and BAO measurements from DESI, are all required to detect the minimal 0.058-eV neutrino mass at more than 3$\sigma$. \textbf{Right:} BFORE-Wide data will set a strong upper limit on the large-scale inflationary gravitational wave signal, and BFORE-Deep will set a strong upper limit on the degree-scale signal.
\label{fig:tau_mnu_degeneracy} }
\end{figure}

One of the primary goals of the next generation ground-based CMB experiments is to measure the sum of the neutrino masses, a major outstanding question in particle physics. The minimal neutrino mass scale is 0.058~eV. As reviewed in the CMB S4 Science Book \cite{CMBS4-ScienceBook}, going beyond 0.200~eV with particle physics lab measurements is a major ongoing effort requiring the development of new technologies and new measurement approaches. In contrast, CMB measurements are already forecast to have the sensitivity to make a high-significance detection in the next decade \cite{abazajian15}. These forecasts show that achieving a detection also requires improved constraints on optical depth to reionization $\tau$ to break the significant degeneracy between neutrino mass and $\tau$ (See Figure~\ref{fig:tau_mnu_degeneracy}). The current best measurement is from \Planck\ data, which yields \mbox{$\tau=0.0550\pm0.0090$} from their SimBaL simulation-based estimator, and \mbox{$\tau=0.0580\pm0.0120$} from their Lollipop cross-spectrum estimator \cite{planck2016-XLVI,planck2016-XLVII}. The CMB-S4 Science Book \cite{CMBS4-ScienceBook} presents forecasting work showing this uncertainty on $\tau$ leaves enough of a degeneracy to preclude a 3$\sigma$ detection of neutrino mass at the minimal scale, even if CMB-S4 (limited to few-degree and smaller angular scales) achieves an all-sky noise level of 1~$\mu$K-arcmin and is combined with DESI-like measurements of large-scale structure. The CMB-S4 CDT Report\cite{CMBS4-CDT-Report} discusses the role a future balloon mission could play in breaking this degeneracy by measuring $\tau$. 

As discussed in earlier forecasting \cite{bryan16} and shown in Figure~\ref{fig:tau_mnu_degeneracy}, BFORE will break this $\tau$--$m_\nu$ degeneracy by precisely measuring the CMB $E$-modes on large angular scales, yielding an independent measurement of $\tau$ that is nearly 3x improved over \Planck. The projected BFORE error on $\tau$ of $\pm 0.0036$ would yield a 3.6$\sigma$ \cite{CMBS4-ScienceBook} or higher detection of neutrino mass if combined with 1~$\mu$K-arcmin CMB lensing data and DESI BAO data. The significance would still be 3$\sigma$ with BFORE, 5~$\mu$K-arcmin CMB lensing data and DESI. The BFORE $\tau$ error is forecast to be within a factor of two of the all-sky CMB cosmic-variance limit of $\sim \pm 0.0020$.

To turn the anticipated BFORE instrument sensitivity and map depth (calculated with a tool based on the one developed\cite{bryan18} for the TolTEC instrument) into a forecast of the sensitivity to optical depth, as well as forecast constraints on neutrino mass and inflationary gravitational waves, we developed an MCMC-based tool. By re-running all of the forecasts on $\tau$, neutrino mass, and gravitational waves shown in Figure~\ref{fig:tau_mnu_degeneracy}, we showed that our MCMC-based tool yields similar results to the publicly-available CMB4CAST tool developed independently \cite{errard16}. CMB4CAST has been used in planning for Simons Observatory and CMB-S4. Our MCMC-based forecasts are slightly conservative, for instance CMB4CAST projects a BFORE error on $\tau$ of $\pm 0.0030$, 17\% better than we forecast with our MCMC code.

\subsection{Foregrounds and Mission Robustness}

We re-ran our forecasts with several different scenarios to verify that the BFORE $\tau$ measurement will be robust to unforeseen contingencies in the mission. Each BFORE EE data point on angular scales larger than $\ell$=30 ($\sim6^\circ$) contributes approximately equal statistical weight to the $\tau$ measurement, so the total error level does not rely heavily on high fidelity reconstruction of the very largest angular scales. To verify this, we ran forecasts assuming the instrument could not reconstruct modes on angular scales larger than $\ell=6$ ($\sim30^\circ$) or $\ell=8$ ($\sim22^\circ$). Compared to the nominal forecast, in these two cases the error on $\tau$ only degrades by 4\% and 19\% respectively. This means that if detector instability or sidelobes reduce the BFORE sensitivity on some of the very large angular scale modes, the error on $\tau$ does not degrade catastrophically. (If the data lacks modes on angular scales larger than $\ell$=30 ($\sim6^\circ$), the error on $\tau$ is forecast to catastrophically degrade by 109\%.) As another test of mission robustness, to determine the impact of suboptimal detector sensitivity and/or reduced detector fabrication yield, we re-ran the forecast with 5\% less instrument sensitivity. We found that the error on $\tau$ only degrades by 1\% due to the measurement being partially cosmic variance limited.

As shown by the map depths in Table~\ref{tab:inst}, BFORE's own data at 217, 280, and 353 GHz will be sufficient to clean our 150 GHz CMB maps from dust contamination, including the complexities of spatially-varying dust spectral index. Existing \Planck\ low frequency data will be sufficient to clean our maps of synchrotron emission. To verify this, we scaled the \Planck\ 30, 44, 70 and 100 GHz map depths \cite{errard16,planck2015-X} by the $\nu^{-3.1}$ synchrotron spectrum to yield equivalent template map depths at 150 GHz of 2.0,~6.7,~18.1,~and~14.8~$\mu$K-arcmin respectively, 
compared to the BFORE 19.2 $\mu$K-arcmin map depth at 150 GHz.
To check this simple scaling for our $\tau$ science goal, we used CMB4CAST \cite{errard16} to forecast that using \Planck\ synchrotron and BFORE dust data to clean the BFORE 150 GHz map would only marginally increase the uncertainty on $\tau$ by 11\% over the foreground-free case.

Ground-based instruments are in an excellent position to take improved large angular scale data at lower frequencies. For example, the CLASS experiment \cite{essinger-hileman/etal:2014,harrington16} will make a synchrotron map at 38 GHz, large angular scale CMB maps at 93 and 145 GHz, and has a single dust channel at 217 GHz. Atmospheric fluctuations may limit CLASS and other ground based experiments in reconstructing large angular scale modes (i.e. larger scales than $\ell$=30, $\theta\sim6^\circ$) at 217 GHz. Also, since no ground-based experiment is currently targeting large angular scale measurements at 280 GHz or higher, ground-based data alone will have a limited ability to account for the spatially-varying dust spectrum or other complications. These factors mean that ground-based experiments may not be able to fully remove dust from 150 GHz maps to precisely measure $\tau$. In contrast, BFORE already has the necessary synchrotron data from Planck, and will use the ULDB mid-latitude balloon platform to access the large angular scales and high frequencies (217, 280, and 353 GHz) needed for robustly cleaning the dust from the BFORE 150 GHz data. Between BFORE and ground based missions, the complimentary overlap in frequency coverage with very different instruments from different environments will provide very strong constraints on systematics in addition to improved sensitivity. Also, the BFORE 280 and 353 GHz maps provide more robust dust foreground removal which will be important in hunting for the large angular scale gravitational wave signal.

\section{Instrument Hardware Status} 

The design of BFORE is driven by the science
goals, which require the large sky coverage uniquely available from the mid-latitude ULDB balloon platform. To get the required instantaneous sensitivity, we selected multichroic detectors with uMUX readout. These detectors use low power readout electronics and the high multiplexing factor reduces cryogenic loading, which in turn reduce the battery weight and liquid cryogen weight. This lets BFORE fit with margin under the ULDB payload weight limit. The detectors and and wide-band anti-reflection coated lenses have high heritage from ACTPol/AdvACT \cite{CMBS4-TechBook}. The readout is based on the system successfully deployed in MUSTANG-2 and baselined for use in the Simons Observatory. 
The telescope, cryostat, and pointed gondola are based on the successful Blast\cite{gandilo14} and Spider \cite{shariff14,soler14} missions. 

\subsection{Detectors}

Microwave multiplexing (uMUX) detector arrays combine the proven performance of TES detectors with microwave multiplexing readout techniques to enable larger detector arrays. Multichroic TES detectors have already been deployed in AdvACT, and similar detectors and the microwave multiplexing are baselined for the Simons Observatory\cite{CMBS4-TechBook}. NIST has already fabricated TES arrays with balloon-suitable saturation power and sensitivity for the second flight of Spider \cite{hubmayr16}. In these proceedings \cite{henderson18,dober18} a mux factor of 512 has been demonstrated.

In lab testing of detectors and readout similar to those we will use in BFORE, $1/f$ noise was not seen in noise spectra taken down to 20 mHz (i.e. 50 seconds) \cite{dober17}. During each night, with the instrument at $45^\circ$ elevation the gondola will completely rotate once every 43 seconds, faster than this measured $1/f$, to reconstruct the large angular scale modes across the half-sky map. Since this rotation corresponds to 6 on-sky degrees per second, 2.6 arcminute beam-scale signals will appear at 138 Hz, well within the 200~Hz detector audio bandwidth, and well-sampled by the 488-Hz readout.

BFORE only requires a modest improvement in per-detector sensitivity over previously flown CMB balloon payloads. This improvement will come mostly from our selection of a Chase compact closed-cycle 100 mK fridge\footnoteref{note1} described in Section 3.3 to cool the detector arrays. Both Boomerang and Spider cooled their detectors with 250 mK fridges.
In its 2003 flight the Boomerang 150 GHz single-polarization detectors had individual sensitivities ranging from 137 to 182 $\mu$K$_\mathrm{CMB}\sqrt{\mathrm{s}}$ \cite{masi06}. The Spider 150 GHz detectors had sensitivities ranging from 140 to 190 $\mu$K$_\mathrm{CMB}\sqrt{\mathrm{s}}$ \cite{rahlin16,gualtieri17} with approximately equal photon noise and thermal fluctuation noise.
Using a 100 mK fridge will reduce the thermal fluctuation noise to enable a sensitivity of 103 $\mu$K$_\mathrm{CMB}\sqrt{\mathrm{s}}$ per detector at 150 GHz. The sensitivity forecasts for the other BFORE bands are presented in Table 1. As discussed in Section 2.2, the BFORE $\tau$ results are forecast to be robust to reduced detector sensitivity or yield.

\subsection{Low Power Detector Readout}

The warm readout electronics consist of an FPGA board capable of processing I/O signals with a total of $>3$~GHz of bandwidth with at least 14 bits of ADC and DAC resolution. Two examples of mSQUID readout systems have already been demonstrated: a ROACH-2 based system (developed by collaborators at NIST) used for the MUSTANG-2 instrument on the GBT and the SMuRF system, consisting of a set of custom boards (developed by SLAC) being developed in the context of the Simons Observatory. The SMuRF system is built for use in ground-based instruments, so the system has not been optimized for low power consumption. Blast-TNG is a balloon mission using KID detectors with similar readout requirements to mSQUID systems, consuming $60$~W with its ROACH-2 based system with 512 MHz bandwidth. For BFORE, we will port the firmware developed for these systems to a new commercial low-power CPU+FPGA system such as the RF SoC ZCU111 board available for purchase from Xilinx. This board has a maximum power consumption of $<25$~W for a single warm readout chain with 4 GHz of bandwidth ($1000-2000$ detectors) compared to $60$~W for the BLAST-TNG ROACH-2 based system with only 512 MHz bandwidth. For reference, the EBEX DfMUX readout electronics had a power consumption of 586~W for 4 crates reading out 954 detectors \cite{Aboobaker} and the Spider TDM readout electronics had a power consumption of approximately 85~W per 1000 detectors.

The reduction in cryogenic loading and warm electronics power consumption simplifies several aspects of instrument design. In balloon projects limited by the weight of liquid helium and solar panels/batteries, it enables substantially longer hold times and the potential to take full advantage of longer flights in the ULDB program.

\subsection{Cryostat and Mission Weight Budget}

The BFORE cryogenic system consists of an optical 1-K cavity inside a long hold-time 500-liter liquid-helium cryostat with a single $<$ 20 cm diameter optical window. The helium is maintained at slightly more than atmospheric pressure during the flight, to minimize loss due to pressure drop at altitude. BFORE uses a continuously operating closed-cycle miniature dilution refrigerator\footnote{\label{note1}Chase Research, Ltd., UK \url{http://www.chasecryogenics.com/Products.htm\#prods3anchor}} This refrigerator does not require any external gas handling, and is buffered by two small closed-cycle Chase $^3$He refrigerators for continuous 100 mK operation. A version of this refrigerator was run continuously for several months at Cardiff. A $^4$He pumped pot cools the optics box to 1\,K.
The detailed design of the BFORE cryostat is adapted from the Blast-TNG 28~day hold-time system. The required LHe volume and cryostat mass can be scaled from the in-flight performance of the Spider cryostat. 
The cryogenic heat load of a system depends chiefly on the input from thermal radiation (dominated by the optical entrance window) and thermal conduction along mechanical supports needed to suspend the cold mass. The BFORE optical system has a relatively small $<$ 20 cm diameter window corresponding to less than 5\% of the Spider total window area. The suspended mass at 4 K in BFORE is approximately 1/4 of the total suspended mass of the 6 Spider optics tubes.
Therefore, roughly scaling the Spider 1200 L helium cryostat by window area or mechanical support area (depending on which dominates the loading) as well as required hold time indicates that BFORE would need approximately (1200 L)$\times$(2x hold time)/(4x less mass or 20x less window area) = 120-600 L of liquid helium. Moving beyond a rough scaling, our detailed cryostat design considers mechanical requirements, window loading, wiring loading, boiloff from cycling the sub-Kelvin fridges, and indicates that we will actually need 500 L to achieve our hold time and other performance requirements. Lab measurements of the boiloff rate of the Blast-TNG cryostat indicate that the design is performing as expected.

In addition to the cryogenic requirements, the BFORE weight and power budget flows down from the science requirements which drive the
detector count needed to achieve the sensitivity at the required frequency bands.
The detector count, combined with our selection of microwave multiplexed cryogenic readout and low-power commercial room temperature readout electronics, in turn flows down to a power consumption requirement. This sets the weight of batteries needed to operate the mission during nighttime observations. These subsystems, combined with weight of all the subsystems needed to point the gondola and handle housekeeping, comprise the weight budget. The total payload weight fits with margin within the current CSBF weight limit, which is expected to increase in the near future as NASA ULDB balloon capabilities continue to mature. This, along with any battery weight reduction enabled by power savings in the readout electronics, will give BFORE additional margin on weight.

\subsection{Stability and Systematics}

Some ground-based experiments use fast polarization modulators to mitigate the impact of detector $1/f$ noise and reduce systematics. This comes at the expense of additional system complexity and potential systematics induced by the modulator itself. The BFORE detectors have been demonstrated to be stable down to frequencies of 20~mHz or better \cite{dober17}
which is stable enough to reconstruct modes on the sky over an entire gondola rotation.
The stability of the detectors and balloon environment means that, as with Spider, rapid polarization modulation is not needed for BFORE.
Spider did require a stepped half-wave plate polarization modulator due to the limited sky rotation at the high latitudes of an Antarctic LDB flight. In contrast, our mid-latitude ULDB flight provides more than 45$^\circ$ of sky rotation, enough to completely switch the detector sensitivity between $Q$ to $U$ without a modulator.

The \Planck\ satellite, in an even more stable environment at L2, was limited at large angular scales by a combination of sensitivity and systematics. The BFORE survey is more sensitive than \Planck\ at all wavelengths and our scan strategy is a more rapid modulation of the sky signal. Both of these factors, along with sky rotation and interleaved $Q/U$ pixels, will enable thorough checks in data analysis for polarimetric and beam systematics. Modern mapmakers such as those used successfully in BICEP/Keck \cite{bicep_III} and ACT \cite{louis16}, use rotation between the sky and instrument to measure and remove beam systematics (such as beam ellipticity and near sidelobes) from true sky signal in post-observation data analysis without a modulator, and BFORE will use the same approach.
Interleaved $Q/U$ pixels have been used successfully for CMB polarimetry even in the ground-based SPTPol instrument, which unlike BFORE, operates at high latitude (at the South Pole) and therefore does not benefit from sky rotation \cite{austermann_spt_2012}.

The balloon environment is more stable than the ground but there are still potential sources of contamination at large angular scales such as differential polarized sidelobe pickup. In BFORE, we minimize potential sidelobes from scattering or diffraction by using a combination of well controlled beams from the horn antennas, a cold optics box, a cold Lyot stop, baffling at the cryogenic window and an extended screen around the primary mirror. The beam sidelobes that persist from the receiver may contribute to near sidelobes if they reflect from the mirror or ground screen, but our approach is designed to minimize far sidelobes since they are particularly challenging to remove in post-observation data analysis.

\section{Conclusions} 

BFORE will use the unique access of mid-latitude ULDB ballooning to large angular scales and high frequencies to address a wide range of CMB science goals. The reionization data will be combined with upcoming ground-based lensing measurements and galaxy surveys to measure neutrino mass with high significance. The B-mode data will detect or set an upper limit on inflationary gravitational waves at both large angular scales and degree scales. The high frequency dust foreground maps will have legacy value for ground-based surveys, and will also be used to measure the CIB and galaxy cluster properties.

The mission is forecast to have margin on sensitivity and detector stability, and existing synchrotron data from Planck is forecast to be sufficient to clean the BFORE maps. New commercial developments in low-power readout electronics enable BFORE to read out a large number of detectors while still fitting in the power and weight budget of a ULDB mission. BFORE leverages developments in detectors, readout, balloon pointing systems, and cryogenics to use the balloon platform to achieve its own science goals, as well as compliment and extend current and future ground-based CMB experiments.
 
\acknowledgments     
 
This work was partially supported by NASA award 80NSSC18K0395.


\bibliography{report}   
\bibliographystyle{spiebib}   

\end{document}